\documentclass[11pt]{llncs}
\usepackage[dvips]{graphicx}
\usepackage{color}

\newcommand{\nop}[1]{}

\newcommand{\mf}[1]{\textcolor{black}{#1}}
\newcommand{\blue}[1]{\textcolor{black}{#1}}
\newcommand{\mm}[1]{\textcolor{black}{#1}}
\title{
Mining Preserving Structures in a Graph Sequence\thanks{
This research is supported by Funding Program for
 World-Leading Innovative R\&D on Science and Technology, Japan.
It is partly supported 
by Grant-in-Aid for Scientific Research (KAKENHI), No.~23500022. 
}
}

\author{
   Takeaki {\normalsize\rm Uno}\inst{1} \and
   Yushi {\normalsize\rm Uno}\inst{2}
}
\institute{
    National Institute of Informatics, 
    2-1-2 Hitotsubashi, Chiyoda-ku, Tokyo 101-8430, Japan. 
    \email{uno@nii.jp}
\and
    Graduate School of Science, Osaka Prefecture University, 
    1-1 Gakuen-cho, Naka-ku, Sakai 599-8531, Japan. 
    \email{uno@mi.s.osakafu-u.ac.jp}
}

\begin{document}

\maketitle

\begin{abstract}
In the recent research of data mining, 
frequent structures in a sequence of graphs have been studied intensively, 
and one of the main concern is changing structures along 
a \mm{sequence of graphs} 
that can capture dynamic properties of data. 
On the contrary, we newly focus on ``preserving structures'' 
in a graph sequence that satisfy a given property for a certain period, 
and mining such structures is studied. 
As for an onset, 
we bring up two structures, a \mm{connected vertex subset} 
and a clique that exist for a certain period. 
We consider the problem of enumerating these structures. 
and present polynomial delay algorithms for the problems. 
Their running time may depend on the size of the representation, 
however, if each edge has at most one time interval in the representation, 
the running time is $O(|V| |E|^3)$ for connected \mm{vertex subsets} 
and $O(\min\{\Delta^5, |E|^2 \Delta\})$ for cliques, 
where the input graph is $G=(V, E)$ with maximum degree $\Delta$. 
To the best of our knowledge, this is the first approach 
to the treatment of this notion, namely, preserving structures. 

\nop{
Mining frequently appearing structures from graph sequences have been 
 studied intensively in the recent data mining area.
Many researches focused on the change of the substructures in the 
 graph sequences, to capture dynamic properties of the data.
On contrary, we focus on preserving structures, that satisfy the
 given properties in certain length.
We consider the problems of enumerating vertex sets that are connected
 graphs (resp., cliques) for certain length.
We propose polynomial delay algorithms for the problems, as well as
 a representation of a graph sequence by the intervals for each edge
 composed of graphs including the edge.
The algorithm for connected graphs is based on recursive partition, 
 and that for cliques is based on reverse search.
Their running time depend on the size of the representation.
In particular, if each edge has at most one interval in the representation,
 the running time is $O(\min\{\Delta^5, |E|^2 \Delta\})$  for cliques,
 and $O(|V| |E|^3)$ for connected components, where the input graph
 is $G=(V, E)$.
To the best of our knowledge, this is the first approach 
 to the treatment of this notion, namely, preserving structures. 
}

\end{abstract}

\section{Introduction}

In a computerized society and 
in the era of explosive growth in data volumes, 
nobody can doubt the importance of data mining, 
that is, extracting useful information (knowledge) 
from a huge data repository. 
A classic \mf{research} of data mining, \mf{for example,} 
is finding association rules 
from a relational database \cite{AIS83}. 
We can classify raw data by its type, e.g., 
numerical data, relational data, structured data, and so on. 
Among these types, data that has a certain kind of {\em graph structure} 
(graph structured data) has become important, 
since it can represent a variety of complex objects 
that appear in practical applications 
such as genome interactions, chemical compounds, hyperlinks on the Web, 
and XML (so-called semi-structured data). 
%


%
Extracting useful facts from graph structured data is often achieved 
by specifying and/or finding frequent substructures 
in a graph. 
In other words, {\em pattern mining} in graphs (or {\em graph mining}) 
\cite{HUS07,IWM03,YH02}. 
In the case of hyperlink structure of the Web (namely, the webgraph), 
for example, 
a clique is considered to be formed by a community focused on a certain topic, 
and finding it may be useful for tracing a social phenomenon on the Web 
\cite{UOU07}. 
These observations imply that one of the most promising approaches 
for graph mining is by {\em enumeration}, 
and efficient enumeration of crucial substructures has a rich history. 
As for cliques, a theoretically efficient algorithm is presented 
in \cite{MkUn04}, and both \cite{MkUn04} and \cite{tomita} 
are state-of-the-art algorithms that performs well in practice. 
Enumerations of paths and matchings are studied in \cite{RdTj75} 
and \cite{FkMt93}, respectively, 
and enumeration of connected components is studied in \cite{AvFk96}. 
Here, we remark that all these algorithms work on a single graph. 

In a practical situation, however, it is often the case that
 graph structures may change over time, and such data is collected
 periodically along a time series. 
In this setting, not only information acquired separately from
 single graphs but also 
from graph patterns appearing sequentially could be important. 
Along this direction, there are some research topics of interest so far. 
Finding graph patterns that appear periodically in a graph sequence 
is studied in \cite{periodic2,periodic1}. 
Graph patterns frequently appear during a certain period are also studied in 
 \cite{k-interval}. 
On the other hand, 
some research address the change patterns 
that appear frequently in a graph sequence composed of graphs 
with edge insertions/deletions, 
such as changes between two time periods \cite{change2} 
and changes of subsequences \cite{change1}. 
Furthermore, 
there are several studies focusing on clustering of vertices 
by utilizing graph sequences 
\cite{seq-graph-cls1,seq-graph-cls3,seq-graph-cls2}. 
We mention here that to achieve these objectives, 
enumeration is again a powerful tool that no other approaches can match. 


\medskip
\noindent
{\bf Objective.} 
Taking these preceding research, 
we propose a new concept of graph mining, 
that is, finding a part of a graph that satisfies a given property 
\mf{continuously} for a long time. 
More specifically, we consider the problem of enumerating all substructures 
 that satisfy a given property during a prescribed period, i.e., 
those appearing in a consecutive subsequence of a graph sequence. 
We name such structures {\em preserving structures} in a graph sequence, 
 and the problem for enumerating all such structures {\em preserving
 structure mining} \mf{in general}. 
As for such properties, 
we consider connected vertex subsets and cliques, in this paper. 
For example, a topic on the Web that is controversial for a long time 
 may correspond to a clique that exists in a consecutive 
 sequence of webgraphs during a certain period. 
As another example, a group of a species in a wildlife environment 
may constitute a consecutive sequence of connected vertex subsets 
 in a sequence of graphs that are constructed from its trajectory
 data \cite{trajectory2,trajectory1}.
To the best of our knowledge, 
 this study is the first case in which a ``long-lasting'' structure 
 is regarded as the target structure to be found. 

\medskip
\noindent
{\bf Related works.} 
(1) {\rm Pattern mining in graph sequences}. 
This is already explained just before the objective. 

(2) {\rm Dynamic flow}. 
On a dynamic network defined by a graph with capacities and transit times 
\mf{along} its edges, the dynamic flow problem asks the maximum flow 
from a specified source to a sink within a given time bound \cite{FF62}. 
As explained later, our model for a graph sequence 
can be naturally generalized so that it implies dynamic flows. 

(3) {\rm Dynamic graph algorithms}. 
Dynamic graph problems concern with a construction of data structures 
 that enables to answer a given graph property quickly, with small 
 update cost for edge insertions/deletions. 
Typical properties of concern include 
connectivity \cite{EpGlItNs92,GHW09}, transitive closures \cite{L11}, 
\mm{cliques \cite{Stix}}, 
bipartiteness, shortest path distance, and so on. 
Dynamic graph algorithms 
could also find a period during which a property is satisfied, however, 
it cannot extract local structures efficiently in a straightforward way. 
For example, we have to find time periods for all possible local structures, 
which may cause exponentially long time for computation. 

\medskip
\noindent
{\bf Contributions.} 
In this paper, we propose a new notion, that is, a preserving structure 
 in a graph sequence.
Then by adopting this newly introduced notion, 
 we pose two problems of mining preserving structures: 
 one for cliques and the other for connected vertex subsets. 
As we have seen so far, both structure or property will play important
 roles in a sequence of graphs that appear in practical situations. 

We then propose efficient algorithms for solving the problems 
by enumerating all vertex subsets 
that are connected or cliques for a certain time period 
in a given graph sequence. 
For this purpose, we define a way of representing a graph sequence 
as the input format. 
In this model, instead of representing a graph at each time 
by the difference from the previous one, 
which is used in the dynamic graph model, 
we represent a graph sequence by explicitly associating each edge 
with its time interval(s) during which it exists. 
Our model is novel and differs from the existing ones (e.g., \cite{binhminh}) 
in the sense that it 
gives a new perspectives for graphs that change over time. 
This graph model introduces a new parameter, 
 namely, the number of time intervals. 
Since the running time of these algorithms could be estimated 
by using this parameter, 
we \mf{also} consider that it would be used as a new measure in the complexity study. 
 
Our algorithm for enumeration of preserving connected vertex subsets 
is based on a recursive graph partition 
and of preserving cliques 
is based on the reverse search, 
which is a framework for designing efficient enumeration algorithms. 
While a straightforward application of maximal clique enumeration 
to our problem requires exponential time, 
our algorithm exploits properties of the time intervals of edges 
so that the algorithm will be polynomial delay. 
Compared to a naive algorithm, this reduces the time complexity 
with a factor of $|E|$. 

\medskip
\noindent
{\bf Organization of this paper.} 
\mf{We first give definitions and representations of graph sequences 
and preserving structures together with basic terminology 
in Section~\ref{Preliminaries}.} 
\mf{In Section~\ref{PreservingConnectedComponent}}, 
we deal with the enumeration problem of preserving connected vertex subsets 
Then we discuss about the closed active clique enumeration problem 
in Section \ref{PreservingClique}. 
We conclude this paper in Section \ref{Conclusion}.

\section{Preliminaries}
\label{Preliminaries}


\subsection{\mf{A Graph Sequence and its Representation}}

A {\em graph} $G$ is an ordered pair of a {\em vertex set} $V$
 and an {\em edge set} $E$, and is denoted by $G=(V,E)$. 
We suppose that \mf{a} vertex set $V$ is $\{1,\ldots,n\}$ so that each vertex has 
 an index and can be treated as an integer.
\mf{The {\it neighborhood} of a vertex $v\in V$ 
is the set $N(v)=\{u\in V\mid \{u,v\}\in E\}$}. 
\mf{The {\it degree} of a vertex $v$ is $|N(v)|$, 
and is denoted by $\deg(v)$. } 
we use $\Delta$ to denote the maximum degree of a graph. 
For a vertex subset \mf{$U$} $(\subseteq V)$, 
the {\em induced subgraph} $G[\mf{U}]$ \mf{of $G$ by $U$} is the subgraph 
whose vertex set is \mf{$U$} and edge set is composed of all edges in $E$ 
that connect vertices in \mf{$U$}. 
\mf{For an edge set $F$, let $V(F)$ denote the set of vertices 
that are endpoints of some edges in $F$}. 
\mf{Then for an edge subset $F$ $(\subseteq E)$, we define 
the induced subgraph $G[F]$ of $G$ by $F$ by the subgraph $G[V(F)]$.}

A {\em time stamp} is an integer representing a discrete time, 
and we denote by ${\cal T}$ 
the ground set 
of all possible {\rm time stamps} during which our graph 
\mf{is} supposed to exist. 
We assume ${\cal T}=\{ 1,\ldots, t_{\max} \}$ without loss of generality, 
and a subset $T$ of ${\cal T}$ is called a {\em time stamp set}. 
We say that an edge of a graph is {\em active} at time stamp $t$ 
if it exists at that moment. 
\mf{The edge set $E$ of a supposed graph consists 
only of edges that are active at some time stamps.} 
To represent a graph sequence, 
we associate a time stamp set with each edge on which it is {\rm active}, 
which we call an {\em active time stamp set} of \mf{that edge}. 
\blue{We regard the active time stamp set of edges 
as a mapping $\tau: E\rightarrow 2^{\cal T}$, 
and thus we define a {\it graph sequence} as a pair of a graph $G=(V,E)$ 
and a mapping $\tau$, 
that is, $(G,\tau)$. }
\mf{Then the active time stamp set of an edge $e$ is $\tau(e)$, 
and we define the {\rm active time stamp set} of an edge set \mf{$F$} 
to be $\tau(F)=\bigcap_{e\in F}\tau(e)$.}
\mf{Given a graph sequence $(G,\tau)$,} 
we define a {\em closure graph} $G_T$ of $G$ 
for a time stamp set $T$ $(\subseteq {\cal T})$ 
\blue{as the spanning subgraph in which its edge set consists of edges 
whose active time stamp sets includes $T$, that is, 
\mf{$G_T=(V,\{e\mid e\in E, T\subseteq \tau(e)\})$.}} 
Especially in case of $T=\{t\}$, a singleton, 
we sometimes denote the closure graph for $T$ by $G_t$ by convention. 
Intuitively, $G_t$ represents a snapshot of $G$ at time stamp $t$. 
By definition, $G_T$ becomes $G$ if $T=\emptyset$.

A time stamp set is ({\em time}) {\em interval} 
if it constitutes \mf{a single interval} $\{t,t+1,\ldots ,t+\ell\}$ 
$(\ell\ge 0)$. 
In this paper, it is sometimes assumed that the active time stamp set 
of any edge is interval, 
and \mf{we call} this an {\em interval assumption}. 
Note that we can assume this without loss of generality, 
since if an active time stamp set of an edge 
is composed of multiple time intervals, 
we can replace it by a set of parallel edges (multi-edge) 
each of which has one of their intervals, respectively. 
%
%
Unlike the existing ones, 
\mf{this way of representing a graph sequence} has an advantage 
in its extendability. 
As for a natural extension, 
we consider edges connecting a vertex 
with a  past time stamp to one with a future time stamp. 
In this case, 
an edge could be represented together with its time interval 
by a tuple of five values $(u,v,t_u,t_v,\ell)$, 
that is, vertices $u$ and $v$ are adjacent by an edge 
from time $t_u$ and $t_v$ until $t_u+\ell$ and $t_v+\ell$, respectively. 
By regarding the pair of a vertex and a time stamp as a kind of super-vertex, 
it can be seen as a set of parallel edges, 
thus we call this extension a {\em thick edge graph}. 

Although this \mf{thick edge graph} model might seem unusual, 
it has several natural applications. 
\mf{One} example is a similarity graph on sequential data.
We regard a pair of a sequence and a time stamp as a vertex.
We draw an edge between two vertices when their corresponding sequences
 at the corresponding time stamps are similar.
In a sequential data, two subsequences are often similar 
in consecutive time intervals, 
 thus thick edges can represent the data in a compact way.
Another example is a dynamic flow, a dynamic version of a network flow. 
In a dynamic flow, when a flow departs a vertex at time $t$ along an
 edge $e$, it arrives at the other end \mf{vertex} of $e$ 
at time $t+\ell(e)$, where $\ell(e)$ is the length of $e$. 
Each edge has its own capacity $c(e)$, and thus pushing flow of $f$
 units along $e$ takes $f/c(e)$ time.
Therefore, a flow along an edge is equivalent to a thick edge. 
Thus, preserving structures in a thick edge graph 
correspond to those composed of thick edges 
that share vertices for sufficiently long time periods. 

\subsection{\mf{Preserving Structures}}

Let $(G,\tau)$ be a graph sequence, 
where $G=(V,E)$ and $\tau: E\rightarrow 2^{\cal T}$ 
with a ground time stamp set ${\cal T}$. 
We next consider preserving structures in a graph sequence, 
that is, a subgraph that consecutively satisfies certain properties, 
such as connected \mm{subgraphs} and cliques in this paper. 
Especially, we will be interested in maximal one of those in some sense. 
We note that the term ``closed'' appearing below 
is employed from the pattern mining field \cite{Pq1}; 
a closed pattern is a maximal pattern that is not included 
in the other patterns with the same frequency.

A vertex subset $U$ is {\it connected} 
if there exists a path between any two vertices of $U$. 
In this case we also say that \mf{$G[U]$} is connected. 
%
A vertex subset 
\blue{$U$ is said to be {\it connected on a time stamp set} $T$ 
if $U$ is connected at any time stamp in $T$,}
and let \mf{$\gamma(U)$ be the set of 
time stamps at which $U$ is connected.} 
We say that a connected vertex subset $U$ is {\em closed} 
if none of its superset $U'$ satisfies $\gamma(U)=\gamma(U')$. 

A {\em clique} is a complete subgraph of a graph.
In this paper, we define a clique by its edge set, 
and thus we do not regard a single vertex as a clique. 
A clique is called {\em maximal} if none of its superset becomes a clique. 
An edge set $F$ is called {\em active} if $\tau(F)\neq \emptyset$, 
and $\tau(F)$ equals ${\cal T}$ if $F=\emptyset$. 
\mf{An active} clique $K$ in a graph sequence is {\em closed} 
if no other clique $K'$ such that $K\subset K'$ satisfies $\tau(K)=\tau(K')$. 
%

\section{Enumeration of Preserving Connected Components}
\label{PreservingConnectedComponent}

In this section we study the closed connected vertex subsets 
in a graph sequence 
\mm{$(G,\tau)$, where $G=(V,E)$ and $\tau: E\rightarrow 2^{\cal T}$ 
with a ground time stamp set ${\cal T}$.} 
We start by observing some properties on closed connected vertex subsets, 
and then present how they can be enumerated. 


\blue{We first have the following simple observations.} 

\begin{property}[closed under union]
\label{union-closed}
For two vertex subsets $U$ and $U'$, 
if both $U$ and $U'$ are connected on a time stamp set $T$ 
and $U\cap U'\neq \emptyset$, then $U\cup U'$ is also connected on $T$. 
\end{property}


\blue{For two partitions ${\cal P}$ and ${\cal P}'$ of a universal set, 
let ${\cal P}\wedge {\cal P}'$
 denote the partition composed of subsets 
given by the intersection of members of 
 ${\cal P}$ and ${\cal P}'$, i.e., ${\cal P}\wedge {\cal P}' 
=\{ I \mid I=H\cap H', H\in {\cal P}, H'\in {\cal P}'\}$. 
A {\em connected component} of $G$ is a maximal vertex subset \mf{$U$} 
such that \mf{$G[U]$} is connected.
%
The set of connected components of $G$ gives a partition of the vertex set, 
and we denoted it by ${\cal C}(G)$.} 
For a time stamp set \mf{$T=\{t_{i_1},\ldots,t_{i_k}\}$}, 
let ${\cal P}(G,T)$ denote $\bigwedge_{j=1}^{k}{\cal C}(G_{t_{i_j}})$, 
which forms a partition of $V$. 
%

\begin{property}[partition]
\label{partition}
\mf{A connected vertex subset $U$ on a time stamp set $T$ 
is included in one of a member (vertex subset) of ${\cal P}(G, T)$.} 
\end{property}


\begin{property}[subdivision]
\label{re-partition}
A connected vertex subset $U$ on a time stamp set $T$, 
where $U$ is included in a vertex set $W$, 
is included in a vertex subset of ${\cal P}(G[W],T)$. 
\end{property}


We denote \mf{the family of all maximal connected vertex subsets} of $G$ 
on a time stamp set $T$ 
by ${\cal C}(G,T)$. 
Property \ref{union-closed} ensures 
that ${\cal C}(G,T)$ becomes a partition of $V$. 
In the subsequent discussions in this subsection, 
\blue{suppose 
that a time stamp set $T$ is interval, and 
let $T_{t,\ell}$ denote an interval time stamp set 
$T_{t,\ell}=\{t,t+1,\ldots ,t+\ell\}$.} 
%
%
\mm{In addition, we assume for simplicity that both ends of any 
interval time stamp set $T_{t,\ell}$ can be examined in $O(1)$ time 
by appropriate pre-process and data structures}. 

Then we have the following two lemmas. 

\begin{lemma}
\label{divide}
\mf{For an interval time stamp set $T_{t,\ell}$ with a fixed time stamp $t$,} 
${\cal C}(G,T_{t,\ell})$ for all $\ell$ $(\ge 0)$ can be computed 
in $O(|V||E|^2)$ time.
\end{lemma}

\proof
We first compute ${\cal C}(G,T_{t,0})={\cal C}(G_{t})$, 
which is simply a family of connected components of $G_t$, in $O(|E|)$ time, 
and then compute each ${\cal C}(G,T_{t,i})$ from ${\cal C}(G,T_{t,i-1})$.
Suppose that we have computed ${\cal C}(G, T_{t,i-1})$.
If $U\in {\cal C}(G,T_{t,i-1})$ is connected on $G_{t+i}$, 
$U$ is connected in $T_{t,i}$, thereby a member of ${\cal C}(G,T_{t,i})$.
If not, from Properties \ref{partition} and \ref{re-partition},
any $U'\in {\cal C}(G,T_{t,i})$ for $U'\subseteq U$ is included in
${\cal C}(G[U],\{t+i\})$.
According to Property \ref{re-partition}, we recursively compute 
${\cal P}(G[U'],T_{t,i})$ for all members $U'$ of ${\cal C}(G[U],\{t+i\})$, 
and repeat this until $U'$ becomes connected on $T_{t,i}$.
In this way, we can compute all members of ${\cal C}(G[U],T_{t,i})$.
The time complexity of this computation is $O(|E|)$ for checking the
connectivity of each $U\in {\cal C}(G, T_{t,i-1})$ at time stamp $t+i$, 
 and $O(p|E|i)$ time for the computation of $\cal P$,  
where $p = |{\cal C}(G, T_i)| - |{\cal C}(G, T_{i-1})|$. 
\mm{Now, without loss of generality, 
since any time stamp appears as either a starting or an ending 
time stamp of an edge, we have $\ell=O(|E|)$.} 
Thus, in total, the computation for all $i$ $(0<i\le \ell)$ 
takes $O(|E|^2)$ time for the
 former, and $O(|V||E|^2)$ time for the latter. 
Therefore the statement holds. 
\qed

\begin{lemma}
\label{irredundant}
Any member $U$ in ${\cal C}(G, T)$ is a closed connected vertex subset 
\mf{of $G$ on an interval time stamp set} $T$. 
\end{lemma}

\proof
From the way of a construction of ${\cal C}(G,T)$, 
no superset of a member of ${\cal C}(G, T)$ is connected on $T$.
It implies that for each $U\in {\cal C}(G,T)$, 
no superset of $U$ is connected in $\gamma(U)$.
This concludes the lemma.
\qed


\medskip
Lemma \ref{irredundant} motivates us to compute ${\cal C}(G, T)$
for all possible \mf{interval time stamp set} $T$ 
to enumerate all closed connected vertex subsets. 
For each time stamp $t$, we compute ${\cal C}(G,T_{t,\ell})$ 
for interval time stamp set $T=\{t,t+1,\ldots,t+\ell\}$ 
for all possible $\ell$. 
From Lemma \ref{divide}, this computation can be done in $O(|V||E|^2)$ time.
Thus we obtain the following theorem, 
\mm{where we use again the fact that $\ell=O(|E|)$}. 

\begin{theorem}
In a graph sequence $(G,\tau)$, 
all closed connected vertex subsets 
can be enumerated in  $O(|V||E|^3)$ time. 
\qed
\end{theorem}


\medskip
The correctness of this algorithm relies only on the above three properties,
 therefore the algorithm can be applied to similar 
 connectivity conditions satisfying these properties, 
such as strong connectivity of a directed graph 
and two-edge connectivity of a graph. 

\begin{theorem}
In a graph sequence $(G,\tau)$ in which $G$ is a directed graph, 
all closed strongly connected vertex subsets 
can be enumerated in $O(|V||E|^3)$ time. 
\qed
\end{theorem}

\begin{theorem}
In a graph sequence $(G,\tau)$, 
 all closed two-edge connected vertex subsets in a graph 
 can be enumerated in $O(|V||E|^3)$ time. 
\qed
\end{theorem}

In the case of two-vertex connectivity, 
 Property \ref{union-closed} holds only when 
 the intersection size of two components is no less than two.
Thus, ${\cal C}(G, T)$ could not be a partition of a vertex set. 
Instead of a vertex set, 
we represent a connected vertex subset by all vertex
 pairs included in the subset. 
Using this representation, when two subsets share at most one vertex, 
 the intersection of their representations is the empty set.
Obviously this representation satisfies the other two properties, 
 thus we have the following theorem.

\begin{theorem}
In a graph sequence $(G,\tau)$, 
all closed two-vertex connected vertex subsets 
can be enumerated in $O(|V|^2|E|^3)$ time. 
\qed
\end{theorem}

\section{Enumeration of Closed Active Cliques}
\label{PreservingClique}

This section discusses about the enumeration of all closed active cliques 
in a graph sequence $(G,\tau)$. 
We first give some additional definitions for further arguments 
and observe some basic properties of closed active cliques. 
After that 
we state a simple output polynomial time algorithm as a warm-up, 
and then we present a more efficient algorithm based on the reverse search 
whose time complexity is much smaller than the simple algorithm. 

For a time stamp set $T$, 
let \mf{$N_T(v)=\{w\mid w\in N(v), T\subseteq \tau(\{v,w\})\}$} 
and $N_T(F)=\bigcap_{v\in V(F)}N_T(v)$ for an edge set $F$, 
that is, 
$N_T(v)$ is the set of vertices adjacent to $v$ at all time stamps in $T$ 
and $N_T(F)$ is the set of vertices adjacent to {\it all} vertices
 in $V(F)$ at any time stamp in $T$. 
For an edge set $F$ and a vertex set $U$, $F\setminus U$ denotes the edge
 set obtained from $F$ by removing all edges incident to some vertices in $U$, 
 and $F\cap U$ denotes $F\setminus (V\setminus U)$. 
For an edge set $F$ and a vertex $v$, let $M(F, v)$ denote the set of edges
 connecting $v$ and a vertex in $V(F)$.
Let $\Gamma(F)$ be the set of vertices $v$ 
such that $\tau(F)\subseteq \tau(M(F,v))$.

Now 
let $F_{\le i}$ be the edge set obtained from $F$ 
by removing edges incident to vertices whose index is greater than $i$. 
By definition, $F_{\le i}$ is empty if $i<1$, and is $F$ if $i\ge n$. 
%
A {\em lexicographic order} on a family of sets 
is a total order defined in such a way that 
a set $F$ is smaller than $F'$ when the smallest element in their symmetric
 difference $F\triangle F'$ belongs to $F$.
\blue{For an active clique $K$ in a graph sequence, 
let $X(K)$ denote the lexicographically smallest closed clique including $K$ 
among all closed cliques \mm{$K'$} such that $\tau(K')=\tau(K)$.} 

\subsection{A Simple Algorithm}

Let $(G,\tau)$ be a graph sequence, 
where $G=(V,E)$ and $\tau: E\rightarrow 2^{\cal T}$ 
with a ground time stamp set ${\cal T}$, 
We first observe a few basic properties of closed active cliques 
in a graph sequence. 
Remember that a clique is defined by an edge set in this paper. 

\begin{lemma}
\label{Xcomp}
For any active clique $K$, 
$X(K)$ can be computed in $O(\min\{ |E|, \Delta^2\})$ time.
\end{lemma}

\proof 
We can obtain 
$X(K)$ by iteratively choosing the minimum vertex $v$ in
 $N_{\tau(K)}(K)$ and adding edges of $M(K, v)$ to $K$,
 until $N_{\tau(K)}(K) = \emptyset$.
$N_{\tau(K)}(K)$ can be computed in $O(\min\{ |E|, \Delta^2\})$ time 
by scanning
 all edges adjacent to some edges in $K$.
When we add $N_{\tau(K)}(K)$ to $K$, $N_{\tau(K)}(K\cup N_{\tau(K)}(K))$ 
can be computed in $O(\deg(v))$ time 
by checking whether $\tau(K)\subseteq \tau(\{u,v\})$ or not 
for each $u\in N_{\tau(K)}(K)$.
Therefore the statement holds.
\qed

\begin{lemma}
\label{clo}
For any time stamp set $T$, any maximal clique $K$ in $G_T$ is closed.
\end{lemma}

\proof
If $K$ is not closed, $G_{\tau(K)}$ includes a clique $K'$ 
such that $K\subset K'$. 
Since $T\subseteq \tau(K)$, 
$T\subseteq \tau(e)$ holds for any edge $e\in K'$. 
This implies that $K'$ is a clique in $G_T$, which contradicts the assumption.
\qed


\medskip
Conversely, it is easy to see 
that any closed active clique $K$ is a maximal clique
 in the graph $G_{\tau(K)}$.
This motivates us to \mm{compute} all maximal cliques in all closure graphs
 of possible active time stamp sets for enumerating all closed active cliques.

\begin{lemma}
\label{simple_cac}
All closed active cliques can be enumerated in $O(|V| |E|^3)$ time
 for each, under the interval assumption.
\end{lemma}

\proof
Under the interval assumption, the active time stamp 
 set of any closed active clique is also an interval.
These active time stamp sets satisfy that the both ends of the interval 
are given by the active
 time sets of some edges, thus their number is bounded by $|E|^2$.
Let $\cal K$ be the family of cliques each of which is a maximal clique
 in a closure graph of some of those active time stamp sets.
Then, from Lemma \ref{clo}, we can see that $|{\cal K}|$ is bounded by
 the product of $|E|^2$ and the number of closed active cliques.
By using the algorithm in \cite{MkUn04}, the maximal cliques can be 
 enumerated in $O(|V|+|E|)$ time for each, and thus the maximal cliques in 
 $\cal K$ can be enumerated in $O((|V|+|E|)|{\cal K}|)$ time.
To check whether an enumerated clique $K$ is closed or not, 
 we compute $X(K)$ in $O(|V|+|E|)$ time.
Since a closed active clique can be a maximal clique of $G_T$ 
for at most $|E|^2$ time stamp sets $T$, 
the closed active cliques can be
 enumerated in $O(|V| |E|^3)$ time for each.
\qed

\subsection{An Efficient Algorithm based on the Reverse Search}

The reverse search is a scheme for constructing enumeration algorithms, 
\mm{and was originally proposed} by Avis and Fukuda \cite{AvFk96} 
for some problems such as enumeration of vertices of a polytope.
The key idea of the reverse search is to \mm{define} an acyclic relation 
among the objects \mm{including the ones} to be enumerated. 
An acyclic relation induces a tree, 
which results in the so-called a parent-child relation, 
and we call the tree a {\it family tree.} 
\mm{Hence enumerating objects is realized by traversing the tree 
according to the parent-child relation to visit all the objects.} 
\mm{In fact, the reverse search algorithm performs a depth-first search 
on the tree induced by the parent-child relation, 
and is implemented by a procedure 
for enumerating all children of a given object.} 
\mm{It starts from the root object that has no parent 
and enumerates its children, 
and then it recursively enumerates children for each child.} 

It is easy to see the correctness of the algorithm; that is, 
the tree induced by the 
 parent-child relation spans all the objects, and the algorithm visits
 all the vertices of the tree by a depth-first search.
When a procedure for enumerating children takes at most $O(A)$ time 
for each child,
 the computation time of the reverse search algorithm is bounded by 
 $O(AN)$, where $N$ is the number of objects to be enumerated. 
Hence, if $A$ is polynomial in terms of the input size, 
\mm{the entire reverse search algorithm takes output polynomial time}. 
In the following, we carefully observe the properties of a graph sequence, 
and prove that enumeration of children can be done in polynomial time. 

Now a more efficient algorithm for enumeration of closed active cliques 
\mm{can be designed} based on the reverse search. 
We start with giving some definitions and fundamental observations. 
The scheme of the reverse search has already been applied 
to enumeration maximal cliques \cite{MkUn04}, 
 and our algorithm for closed active cliques adopts their ideas. 
For an active clique $K$, let $i(K)$ be the minimum vertex $i$ satisfying 
 ${X}(K_{\le i}) = K$.
We define the parent $P(K)$ of closed active clique $K$ by
 $X(K_{\le i(K)-1})$, and 
$P(K)$ is not defined for $K = X(\emptyset)$, which is called the {\em root} 
\mm{of the family tree.} 

\begin{lemma}
\label{acyclic}
The parent-child relation defined by $P$ is acyclic. 
\end{lemma}

\proof
Suppose that $K$ is a closed active clique such that $P(K)$ is defined. 
$P(K)$ is generated by removing vertices one by one from $K$, 
 and adding vertices so that the active time set does not change, 
 thus $\tau(P(K))$ always includes $\tau(K)$.
Since $X(K_{\le i(K)-1}) \ne K$, $P(K)$ is lexicographically smaller than 
 $K$ when $\tau(P(K)) = \tau(K)$.
Thus, either (a) $P(K)$ has a larger active time set than $K$, 
or (b) $P(K)$ has the same active time set as $K$ 
and is lexicographically smaller than $K$.
Therefore the statement holds.
\qed
 
\begin{lemma}
\label{ppc}
Any vertex in $P(K)_{\le i(K)} \setminus K$ does not belong to 
 $N_{\tau(K)}(i(K))$, 
and therefore $K_{\le i(K)-1}$ $=P(K)_{\le i(K)}$ $\cap$ $N_{\tau(K)}(i(K))$. 
\end{lemma}

\proof
Suppose that a vertex $v$ in $P(K)_{\le i(K)} \setminus K$ belongs to 
 $N_{\tau(K)}(i(K))$.
Then, $X(K_{\le i(K)})$ has to include either $v$ or another vertex $u < v$.
It implies that $X(K_{\le i(K)}) \cap \{1,\ldots, i(K)\} \ne K_{\le i(K)}$,
 thereby $X(K_{\le i(K)}) \ne K$.
This contradicts the definition of $i(K)$. 
\qed


\medskip
A subset $F$ of $M(K, v)$ is called {\em time maximal} if $F$ is included in
 no other subset $F'$ of $M(K, v)$ satisfying 
$\tau(F)\cap \tau(K) = \tau(F')\cap \tau(K)$.
Let $I(K, v)$ be the set of all time maximal subsets of $M(K, v)$.
For a time maximal subset $F\in I(K, v)$, we define 
 $C(K, F) = X(K_{\le v}\cap V(F)\cup F)$.

\begin{lemma}
\label{child}
If $K'$ is a child of non-root closed active clique $K$,
then $K' = C(K, F)$ holds for some vertex $v$ and $F\in I(K, i(K'))$.
\end{lemma}

\proof
Let $F = M(K_{\le i(K')}, i(K'))$.
From Lemma \ref{ppc}, $K'_{\le i(K')-1} = K_{\le i(K')-1}\cap
 V(N_{\tau(K')}(i(K')))$ holds, 
and thus $K = X(K_{\le i(K')-1}\cap V(F)\cup F)$.

We next show that $F$ is a member of $I(K, i(K'))$.
Suppose that $K'$ is a child of $K$, and $F$ does not belong to
 $I(K, i(K'))$, i.e., $F$ is properly
 included in an edge subset $F'\in I(K, i(K'))$ 
such that $\tau(F) = \tau(F')$.
Then, the active time set of $K_{\le i(K')}\cap V(F')$ is same as 
 that of $K'_{\le i(K')} = K_{\le i(K')-1}\cap V(F)\cup F$.
This implies that $X(K'_{\le i(K')})$ includes several edges in $F'$, 
which contradicts to the definition of $i(K')$.
\qed

%

\medskip
Since $X(K_{\le v})\ne K$ holds for any $v < i(K)$, we have the
 following corollary.

\begin{corollary}
$C(K, F)$ is not a child of $K$ for any $F\in I(K, v)$ satisfying $v < i(K)$.
\end{corollary}

It is true that any child is $C(K, F)$ for some $F$. 
However, $C(K, F)$ cannot always be a child, that is, 
$C(K, F)$ is a child of $K$ if and only if $P(K) = P(C(K, F))$. 
This implies that 
 we can check whether $C(K, F)$ is a child or not by computing $P(K)$.
Therefore, from Lemma \ref{child}, we obtain the following \mm{procedure} 
 to enumerate children of $K$.
For avoiding the duplicated output of the same child $K'$, we output $K'$
 only when $K'$ is generated from $F\in I(K, i(K'))$.

\renewcommand{\arraystretch}{0.95}
\begin{center}
\begin{tabular}{l}\hline 
{\bf Procedure} EnumChildren($K$: non-root closed active clique) \\ \hline
1. {\bf for each } $F\in I(K, v), v>i(K)$ {\bf do}\\
2. \ \ \ \ compute $C(K, F)$; \\
3. \ \ \ \ compute $i(C(K, F))$ and $P(C(K, F))$; \\
4. \ \ \ \ {\bf if } $K = P(C(K, F))$ and $i(C(K, F)) = v$ {\bf then}
    {\bf output} $C(K, F)$; \\
5. {\bf end for} \\ \hline 
\end{tabular}
\end{center}


For analyzing the complexity of this \mm{procedure}, 
which will later be used as a subroutine of the entire algorithm 
for enumerating closed active cliques, 
we show some technical lemmas. 

\begin{lemma}
\label{Pcomp}
$P(K)$ can be computed in $O(|E|)$ time.
\end{lemma}

\proof
Suppose that $K$ is not the root, i.e., $P(K)$ is defined. 
Let $K'$ be initialized to the empty set, and we add vertices of $K$
 to $K'$ one by one from the smallest vertices in the increasing order.
In each addition, we maintain the change of $\tau(K')$ and $N_{\tau(K)}(K')$. 
Then, we can find the minimum vertex $v$ satisfying $\tau(K_{\le v})=\tau(K)$, 
 and the minimum vertex $u$ satisfying
 $i = \min \{N_{\tau(K)}(K_{\le i-1}) \}$ for any $i\in K, i\ge u$.
We have $i(K) = \max \{u, v\}$, since $X(K_{\le j}) \ne K$ holds when either
 $\tau(K)\ne \tau(K_{\le j})$ 
or $i \ne \min\{N_{\tau(K)}(K_{\le i-1}) \}$ holds
 for some $i\in K, i>j$.
Under the assumption that both ends of any interval time stamp set 
can be examined in $O(1)$ time, 
$\tau(K\cup \{ e\})$ can be computed in 
 $O(1)$ time from $\tau(K)$ for any edge $e$.
Thus, we can compute $i(K)$ in $O(\min \{|E|, \Delta^2\})$ time.
Together with Lemma \ref{Xcomp}, the statement holds.
\qed

\begin{lemma}
\label{non-root-child}
If $K$ is not the root, any child $K'$ of $K$ satisfies that
 $K_{\le i(K')}\cap K'_{\le i(K')}\ne \emptyset$.
\end{lemma}

\proof
If $K_{\le i(K')}\cap K'_{\le i(K')}= \emptyset$,
 it holds that $K'_{\le i(K')-1}\cap K = \emptyset$.
Since $K'_{\le i(K')-1}$ is always included in $K$, we have 
 $K'_{\le i(K')-1} = \emptyset$.
Therefore, $P(K') = X(\emptyset)$, which implies that $P(K)$ is the root. 
\qed

\begin{lemma}
\label{numchild}
If $K$ is not the root, the children of $K$ is enumerated by evaluating
 at most $\min \{ \Delta|E|, \Delta^3\}$ edge sets under the
 interval assumption.
\end{lemma}

\proof
By the interval assumption, the ends of the active time set of
 any subset $F$ of $I(K, v)$ is given by the ends of some edges in $F$, 
and thus $|I(K, v)|$ is bounded \mm{from above} by $\Delta^2$. 
Lemma \ref{non-root-child} ensures that if $K$ is not the root,
 \mm{Step 2} of EnumChildren does not have to take care of vertices
 not adjacent to any vertex of $V(K)$.
This means that we have to take care only of non-empty maximal subset in
 $I(K, v)$.
Let $I$ be the union of all non-empty subsets of $I(K, v)$.
Since each edge in $F\in I(K, v)$ is incident to some vertices in $K$, 
 we have $|I| \le \min\{ |E|, \Delta^2\}$.
It implies that the number of possible choices of two edges from 
 some non-empty $I(K, v)$ is bounded 
\mm{from above} by $\Delta\cdot\min\{ |E|, \Delta^2\}$.
\qed


\medskip
\mm{By the above lemmas, 
we can estimate the time complexity of the procedure of enumerating children.} 

\begin{lemma}
\label{enumchild}
\mm{Procedure} EnumChildren enumerates all children of $K$ in
 $O(\min \{ \Delta^5, |E|^2\Delta\})$ time under the interval assumption.
\end{lemma}

\proof 
The correctness of the \mm{procedure} comes from Lemma \ref{child}. 
We note that the \mm{procedure} never output any child more than once, 
 since each child is generated from \mm{its unique parent, 
a maximal subset} 
 included in $F\in I(K, i(K))$.
We then observe that all non-empty subset $F\in I(K, v), v>i(K)$ can be
 computed in $O(\min\{|E|, \Delta^2\})$ time by scanning all edges
 adjacent to some edges in $K$, and $C(K, F)$ can be computed in
 $O(\min \{ |E|, \Delta^2\})$ time in a straightforward manner.
From Lemma \ref{numchild}, the \mm{procedure iterates} the loop for
 $\min \{ \Delta|E|, \Delta^3\}$ edge sets, and each edge set \mm{spends} 
 $O(\min \{ |E|, \Delta^2\})$ time from Lemma \ref{Pcomp}.
Thus, we conclude the lemma.
\qed

\medskip
Now we describe our algorithm 
\mm{for enumerating all closed active cliques in a graph sequence 
based on the reverse search as follows. 
It is presented in a slightly different form 
by introducing a threshold $\sigma$ with respect to the length 
of active time stamp sets 
by observing that $\tau(K)\subseteq \tau(P(K))$ always holds. 
It enumerates all closed active cliques having active time sets 
 larger than $\sigma$ by giving $X(\emptyset)$ 
(thus enumerates all when $\sigma$ is set to be $0$)}. 

\renewcommand{\arraystretch}{0.95}
\begin{center}
\begin{tabular}{l}\hline 
{\bf Algorithm} EnumClosedActiveClique($K$: closed active clique)\\ \hline
\ \ 1. {\bf output } $K$; $prv := nil$; \\
\ \ 2. {\bf if} $prv = nil$ {\bf then} $K' :=$ the first clique found by EnumChildren($K$); \\
\ \ \ \ \ \ {\bf else} $K' := $ the clique found just after $prv$ by EnumChildren($K$); \\
\ \ 3. {\bf if} there is no such clique $K'$ {\bf go to} \mm{Step 8}; \\
\ \ 4. $K := K'$; free up the memory for $K'$; \\
\ \ 5. {\bf if} $|\mm{P(K)}|\ge \sigma$ {\bf then call} EnumClosedActiveClique($K$); \\
\ \ 6. $K := P(K)$; \\
\ \ 7. {\bf go to} \mm{Step 2}; \\
\ \ 8. {\bf if} $K$ is not the root {\bf then} return; \\
\ \ 9. {\bf for each} $e\in E$ {\bf do}\\
10. \ \ \ {\bf if} $e$ is lexicographically minimum in $X(e)$ {\bf then}
 EnumClosedActiveClique($X(e)$); \\
11. {\bf end for} \\ \hline
\end{tabular}
\end{center}


\medskip
Finally, we can establish the following theorem. 

\begin{theorem}
\label{ecac}
Under the interval assumption, 
\mm{Algorithm EnumClosedActiveClique enumerates} 
all closed active cliques in a graph sequence 
in $O( N\min \{ \Delta^5, |E|^2\Delta\})$ time 
\mm{and in $O(|V|+|E|)$ space,} 
where $N$ is the number of closed active cliques in a graph sequence. 
\end{theorem}

\proof
The correctness of the algorithm is easy to see from the framework of 
the reverse search and Lemma \ref{acyclic}.
The computation time of the reverse search is given by the product of the
 number of objects to be enumerated and the computation time on each object.
From Lemma \ref{enumchild}, an iteration \mm{requires} 
 $O(N\min \{ \Delta^5, |E|^2\Delta\})$ time for non-root closed active
 cliques.
For the root $K=X(\emptyset)$, we can enumerate its children $K'$
 satisfying the condition of Lemma \ref{non-root-child} in 
 $O(N\min \{ \Delta^5, |E|^2\Delta\})$ time using \mm{procedure} EnumChildren.
When $K_{\le i(K')}\cap K'_{\le i(K')}= \emptyset$, 
 we have $K'_{\le i(K')-1}\cap K = \emptyset$.
This implies that $K'_{\le i(K')}$ is composed of an edge, thus 
 by generating $X(\{ e\})$ for all $e\in E$, we can enumerate the children
 that do not satisfy the condition of Lemma \ref{non-root-child}, in 
 $O(\min \{|E|^2, |E|\Delta^2\})$ time.
Note that the duplication can be avoided by outputting $X(\{ e\})$
 only when $e = \arg \min X(\{e\})$.
Since $N \ge |E|/\Delta^2$, it holds that $\min \{|E|^2, |E|\Delta^2\} \le
 N\min \{ \Delta^5, |E|^2\Delta\}$.
Therefore the time complexity of the algorithm is as stated.

In a straightforward implementation of the algorithm, each iteration
 may take $\omega(|V|+|E|)$ space for keeping the intermediate results of
 the computation in memory, especially for all $C(K,F)$. 
We can reduce this by restarting the iteration from the beginning.
When we find a child $K'$ \mm{of $K$}, 
we immediately generate the recursive call 
 with $K'$, before the termination of the enumeration of the children.
After the termination of the recursive call, we resume the enumeration 
 of the children.
To save the memory, we restart from the beginning of the iteration,
 and we pass through the children found before $K'$, and reconstruct all 
 the necessary variables.
We note that the time complexity does not change by the restart, 
 since the number of restarts is bounded by the number of recursive
  calls generated by the algorithm.
A child is given by a maximal edge subset, and a maximal edge subset
 is given by two edges.
Thus, we can memorize a child by a constant number of variables.
The clique $K$ is constructed by computing $P(K')$, thus 
\mm{it is also not necessary to have $K$} 
in memory, and can be re-constructed 
 without increasing the time complexity.
The iteration with respect to the root takes $O(|V|+|E|)$ \mm{space}, 
\mm{therefore we have the atatement of the theorem.} 
\qed

\medskip
\mm{As we stated,} since $\tau(K)\subseteq \tau(P(K))$ always holds, 
we have the following corollary.

\begin{corollary}
Under the interval assumption, 
\mm{Algorithm EnumClosedActiveClique enumerates} 
all closed active cliques
 having active time sets no shorter than a given threshold $\sigma$ 
 in $O(\min \{ \Delta^5, |E|^2\Delta\})$ time 
 for each \mm{and in $O(|V|+|E|)$ space}. 
\qed
\end{corollary}

Note again that the interval assumption can be set 
without loss of generality, 
since 
 we can replace an edge with multiple time intervals 
by parallel edges having a single time interval for each, 
in their active time stamp sets.
However, this transformation increases the degrees of the vertices, 
 thus the time complexity may increase. 
If we set $\Delta$ to the maximum degree to the transformed graph, 
then the results hold. 

\vspace*{-0.2cm}
\subsection{Extension to \mm{Thick Edge Graphs}}

\vspace*{-0.1cm}
We consider the extension of our algorithm to ``thick edge graphs''. 
In a thick edge graph, a clique is composed of several vertices 
 with different time stamps.
Hence, a clique is supposed to be 
``vertex $v_1$ at time stamp $t_1$, 
$\ldots$ , and vertex $v_k$ at time stamp $t_k$ are fully connected''. 
We thus associate a non-negative number {\em shift} $s(v)$ for each
 vertex $v$ to define the active time stamp set for vertex sets. 
For an edge set $K$ and a set $S$ of shifts $s(v)$ for vertices $v$ in
 $V(K)\mm{= \{v_1,\ldots,v_k\}}$, 
their active time stamp set is defined by the set of $t$ such that 
 ``vertex $v_1$ at time stamp $s(v_1)+t$, $\ldots$ , and vertex $v_k$ 
at time stamp 
  $s(v_k)+t$ form a clique''. 
We exclude its ambiguity by setting one of $s(v)$ to 0.

A closed active clique in a thick edge graph is defined by a pair of an edge
 set $K$ and shifts $S$ such that no clique with the same shift
 for vertices of $V(K)$ includes $K$.
Once we fix shifts of all vertices in the graph, the enumeration of 
 closed active cliques in a thick edge graph is equivalent to that 
 in a graph sequence. 
Although the exhaust search may take exponential time, 
our enumeration algorithm based on the reverse search still works 
even in thick edge graphs. 

First, we define the lexicographic order on the set of pairs of a vertex
 and \mm{its} shift, i.e., $\{ (v_1,s(v_1))$, $\ldots$, $(v_k,s(v_k))\}$. 
Then, $X(K)$ and $P(K)$ are defined in the same way as on a graph sequence, 
 and their computation can be done in the same time complexity.
A child is obtained from its parent by adding a vertex \mm{$w$} and setting 
 the shift of \mm{$w$}, and Lemmas \ref{ppc}, \ref{child} and
 \ref{non-root-child} also hold.
Since the choice of the shift of \mm{$w$} depends on the choice of the edge
 to be added to $K$, the number of children of a closed active clique 
 is also bounded by $\min \{ \Delta|E|, \Delta^3\}$, 
which implies that Lemma~\ref{numchild} also holds.
Thus, we have the following corollary.

\begin{corollary}
Under the interval assumption, all closed active cliques in a 
 thick edge graph, with active time stamp sets no shorter than
 a given threshold $\sigma$ can be enumerated in
 $O(\min \{ \Delta^5, |E|^2\Delta\})$
 time for each within $O(|V|+|E|)$ \mm{space}. 
\qed
\end{corollary}

\vspace*{-0.2cm}
\section{Conclusion}
\label{Conclusion}

\vspace*{-0.1cm}
In this paper, we focused on the structures preserved in a sequence of graphs 
 continuously for a long time, 
which we call ``preserving structures''. 
We considered two structures, closed connected \mm{vertex subsets} 
and closed active
 cliques, and proposed efficient algorithms for enumerating these 
 structures preserved during a period no shorter 
than \mm{a prescribed length.} 
An interesting future work is to develop efficient algorithms for 
preserving structure mining problems for other graph properties.

\nop{
\vspace*{-0.2cm}
\section*{Acknowledgment}

This research is supported by Funding Program for
 World-Leading Innovative R\&D on Science and Technology, Japan.
It is partly supported 
by Grant-in-Aid for Scientific Research (KAKENHI), No.~23500022. 
}

\nop{

}

\thispagestyle{empty}


\begin{thebibliography}{99}
%
\bibitem{AIS83}
R.~Agrawal, T.~Imielinski and A.~Swami.
Mining association rules between sets of items in large databases.
{\it Proc. Int'l Conf. on Management of Data}, 
pp.~207--216 (1993).
%
%
\bibitem{HUS07}
H.~Arimura, T.~Uno and S.~Shimozono.
Time and space efficient discovery of maximal geometric graphs.
{\it Discovery Science}, pp.~42--55 (2007).

\bibitem{AvFk96} D.~Avis and K.~Fukuda.
   Reverse search for enumeration.
   {\it Discr. Appl. Math.}, 65, pp.~21--46 (1996).
%
%
\bibitem{change2}
M.~Berlingerio, F.~Bonchi, B.~Bringmann and A.~Gionis.
  Mining graph evolution rules.
  {\it Lecture Notes in Computer Science}, Vol.~5781, pp.~115--130 (2009).
%
\bibitem{binhminh}
B.~Bui Xuan, A.~Ferreira and A.~Jarry. 
Computing shortest, fastest, and foremost journeys in dynamic networks. 
{\it Int. J. of Foundations of Computer Science}, 14, pp.~267--285 (2003). 

\bibitem{k-interval} K.~M.~Borgwardt, H.~P.~Kriegel and P.~Wackersreuther.
  Pattern mining in frequent dynamic subgraphs.
  {\it Proc. 6th IEEE ICDM},
pp.~818--822 (2006).
%
\bibitem{EpGlItNs92} D.~Eppstein, Z.~Galil, G.~F.~Italiano and
A.~Nissenzweig.
   Sparsification---A technique for speeding up dynamic graph algorithms.
   {\it J. ACM}, 44, pp.~669--696 (1997).
%
\bibitem{FF62}
L.~R.~Ford and D.~R.~Fulkerson.
{\it Flows in Networks}.
Princeton University Press (1962).
%
\bibitem{FkMt93} K.~Fukuda and T.~Matsui.
   Finding all the perfect matchings in bipartite graphs.
   {\it Applied Mathematics Letters}, 7, pp.~15--18 (1994).
%
\bibitem{GHW09}
R.~G\"{o}rke, T.~Hartmann and D.~Wagner.
\newblock Dynamic graph clustering using minimum-cut trees.
\newblock {\it Lecture Notes in Computer Science}, Vol.~5664,
339--350, Springer (2009).
%
\bibitem{periodic2}
J.~Han and G.~Dong and Y.~Yin.
  Efficient mining of partial periodic patterns in time series database.
  {\it Proc. 15th IEEE ICDE},
  pp.~106--115 (1999).
%
\bibitem{change1} A.~Inokuchi and T.~Washio.
  A fast method to mine frequent subsequences from graph sequence data.
  {\it Proc. 8th IEEE ICDM},
  pp.~303--312 (2008).
%
\bibitem{IWM03}
A.~Inokuchi T.~Washio and H.~Motoda.
Complete mining of frequent patterns from graphs: Mining graph data.
{\em Machine Learning}, 50, pp.~321--354 (2003).

\bibitem{trajectory2} 
P.~Kalnis, N.~Mamoulis and S.~Bakiras. 
    On discovering moving clusters in spatio-temporal data. 
    {\it Proc. 9th SSTD}, pp.~364--381 (2005). 
%
\bibitem{L11}
J.~\L\c{a}cki.
Improved deterministic algorithms for decremental transitive closure
and strongly connected components.
{\it Proc. 22nd ACM-SIAM SODA},
pp.~1438--1445 (2011).

\bibitem{periodic1}
M.~Lahiri and T.~Y.~Berger-Wolf.
  Mining periodic behavior in dynamic social networks.
  {\it Proc. 8th IEEE ICDM},
   pp.~373--382 (2008).
%
\bibitem{trajectory1} 
Z.~Li, B.~Ding, J.~Han and R.~Kays. 
    Swarm: Mining relaxed temporal moving object clusters. 
    {\it Proc. 36th Int'l Conf. on VLDB}, 
      pp.~723--734 (2010).
%
\bibitem{MkUn04} K.~Makino and T.~Uno.
   New algorithms for enumerating all maximal cliques.
   {\it Lecture Notes in Computer Science}, Vol.~3111, pp.~260--272 (2004).
%
\bibitem{Pq1} N.~Pasquier, Y.~Bastide, R.~Taouil and L.~Lakhal.
  Efficient mining of association rules using closed itemset lattices.
  {\it J. Information Systems}, 24, pp.~25--46 (1999).
%
\bibitem{RdTj75} R.~C.~Read and R.~E.~Tarjan.
   Bounds on backtrack algorithms for listing cycles, paths,
   and spanning trees.
   {\it Networks}, 5, pp.~237--252 (1975).
%
\bibitem{Stix}
\mm{V.~Stix. 
Finding all maximal cliques in dynamic graphs. 
{\it Computational Optimization and Applications}, 27, pp.~173--186 (2004).} 
%
\bibitem{seq-graph-cls1}
J.~Sun, C.~Faloutsos, S.~Papadimitriou and P.~S.~Yu.
  GraphScope: Parameter-free mining of large time-evolving graphs.
  {\it Proc. 13th ACM Int'l Conf. on KDD},
  pp.~687--696 (2007).
%
\bibitem{seq-graph-cls3}
C.~Tantipathananandh and T.~Berger-Wolf.
  Constant-factor approximation algorithms for identifying dynamic
  communities.
  {\it Proc. 15th ACM Int'l Conf. on KDD},
  pp.~827-836 (2009).

\bibitem{seq-graph-cls2}
C.~Tantipathananandh, T.~Berger-Wolf and D.~Kempe.
  A framework for community identification in dynamic social networks.
  {\it Proc. 13th ACM Int'l Conf. on KDD},
  pp.~717--726 (2007).

\bibitem{tomita} E.~Tomita, A.~Tanaka and H.~Takahashi.
  The worst-case time complexity for generating all maximal cliques and
  computational experiments.
  {\it Theor. Comp. Sci.}, 363, pp.~28--42 (2006).
%
\bibitem{UOU07}
Y.~Uno, Y.~Ota and A.~Uemichi.
Web structure mining by isolated cliques.
{\it IEICE Transactions on Information and Systems},
Vol.~E90-D, pp.~1998--2006 (2007).

\bibitem{YH02}
X.~Yan and J.~Han.
gspan: Graph-based substructure pattern mining.
  {\it Proc. 2nd IEEE ICDM},
  pp.~721--724 (2002).

\end{thebibliography}
\end{document}